\documentclass[11pt,a4paper]{article}
\usepackage[utf8]{inputenc}
\usepackage[T1]{fontenc}
\usepackage{geometry}
\usepackage{amsmath}
\usepackage{lineno}
\usepackage{amsmath}
\usepackage{nccmath}
\usepackage{graphicx}
\usepackage{multirow}
\usepackage{makecell}
\usepackage{url}
\usepackage{rotating}
\usepackage{floatrow}
\usepackage{multicol}
\usepackage[onehalfspacing]{setspace}

\usepackage{pgfplots}
\pgfplotsset{compat=1.18}

\usepackage{sfmath}
\usepackage{amssymb}
\usepackage{booktabs}
\usepackage{placeins}
\usepackage{caption}
\usepackage{verbatim}
\usepackage{amsthm}
\usepackage{numprint}

\usepackage{lipsum}
\usepackage[caption=false]{subfig}

\usepackage{titlesec}
\titleclass{\subsubsubsection}{straight}[\subsection]
\newcounter{subsubsubsection}[subsubsection]
\renewcommand\thesubsubsubsection{\thesubsubsection.\arabic{subsubsubsection}}
\titleformat{\subsubsubsection}{\normalsize}{\thesubsubsubsection}{1em}{}
\titlespacing*{\subsubsubsection}{0pt}{3.25ex plus 1ex minus .2ex}{1.5ex plus .2ex}
\makeatletter
\def\toclevel@subsubsubsection{4}
\def\l@subsubsubsection{\@dottedtocline{4}{7em}{4em}}
\makeatother
\newlength{\thickarrayrulewidth}
\def\thickhline{%
  \noalign{\ifnum0=`}\fi\hrule \@height \thickarrayrulewidth \futurelet
   \reserved@a\@xthickhline}
\def\@xthickhline{\ifx\reserved@a\thickhline
               \vskip\doublerulesep
               \vskip-\thickarrayrulewidth
             \fi
      \ifnum0=`{\fi}}
\makeatother

\modulolinenumbers[5]
\theoremstyle{definition}

\usepackage{hyperref}
\hypersetup{
    colorlinks=true,
    linkcolor=blue,
    citecolor=blue,
    urlcolor=blue
}

\begin{document}

\title{Efficient Automated Diagnosis of Retinopathy of Prematurity by Customize CNN Models}

\author{Farzan Saeedi\\
        \small Department of Computer Science, Sultan Qaboos University, Muscat, Oman.\\
        Sanaz Keshvari\\
        \small Department of Electrical and Computer Engineering, Sultan Qaboos University, Muscat, Oman.\\
        Nasser Shoeibi\\
        \small Eye Research Center, Mashhad University of Medical Sciences, Mashhad, Iran}
\maketitle

\begin{abstract}


This paper encompasses an in-depth examination of Retinopathy of Prematurity (ROP) diagnosis, employing advanced deep learning methodologies. Our focus centers on refining and evaluating CNN-based approaches for precise and efficient ROP detection. We navigate the complexities of dataset curation, preprocessing strategies, and model architecture, aligning with research objectives encompassing model effectiveness, computational cost analysis, and time complexity assessment. Results underscore the supremacy of tailored CNN models over pre-trained counterparts, evident in heightened accuracy and F1-scores. Implementation of a voting system further enhances performance. Additionally, our study reveals the potential of the proposed customized CNN model to alleviate computational burdens associated with deep neural networks. Furthermore, we showcase the feasibility of deploying these models within dedicated software and hardware configurations, highlighting their utility as valuable diagnostic aids in clinical settings. In summary, our discourse significantly contributes to ROP diagnosis, unveiling the efficacy of deep learning models in enhancing diagnostic precision and efficiency.

\end{abstract} 


\section{Introduction}

Among the numerous challenges faced by infants with low birth weight and premature birth, Retinopathy of Prematurity (ROP) stands out as a significant threat to their visual health \cite{dammann2023retinopathy}. The timely diagnosis and treatment of ROP can be instrumental in preventing lifelong visual impairment. However, the scarcity of specialized ophthalmologists and the need for immediate attention create a pressing need for alternative solutions \cite{tan2022metaverse}. This article delves into the realm of machine learning and explores the potential of supervised learning algorithms in efficiently detecting ROP using real-world datasets of retina images, shedding light on the promising role of technology in improving access to accurate diagnoses and facilitating timely interventions.

In recent years, the success of deep learning algorithms in image classification tasks has been remarkable \cite{shafiq2022deep}, with convolutional neural networks (CNNs) being the most widely used in the field \cite{wu2023applications}. However, a major limitation of these artificial DNNs lies in their heavy reliance on large-scale training data . Unlike humans who possess inherent knowledge and inference capabilities, artificial DNNs heavily rely on data analysis for understanding. Therefore, when the model is trained on a limited dataset, it may suffer from overfitting, where it becomes too specific to the training data and fails to generalize well to new instances . To mitigate this issue, researchers have employed data augmentation techniques to artificially generate additional similar data, enhancing the model's ability to learn and generalize effectively.

Another significant challenge in the context of ROP is associated with the limitations of the imaging system used. The minimum capability of this tool allows for the coverage of only 170 degrees out of the complete 200 degrees of the eye. To address this issue, experts have the ability to adjust the camera direction and capture the region that was missed in the initial image. In this research, we propose the utilization of image processing techniques to extract the necessary portions from these images. Subsequently, a voting system is employed to aggregate predictions from multiple images, which are utilized for classification. The deep neural network is responsible for performing the actual classification based on these aggregated predictions.

Our research paper makes significant contributions to the field of retinopathy of prematurity (ROP) detection. These contributions are as follows:

\begin{itemize}
    \item Data Augmentation for Limited Real-World Data: Due to the limited availability of real-world data for ROP detection, we address this challenge by employing data augmentation techniques. By artificially expanding the dataset through techniques such as rotation, flipping, and scaling, we create additional training samples. This augmentation process enhances the diversity and variability of the dataset, enabling our model to learn and generalize better to unseen retinal images of premature infants.

    \item Customized CNN Model Design: Recognizing the unique characteristics and challenges associated with ROP detection, we propose a novel CNN architecture specifically designed for this task. Our model incorporates specialized layers, such as convolutional layers, to effectively capture the relevant features and patterns in retinal images. Additionally, we optimize the network architecture by fine-tuning hyperparameters and adjusting the depth and width of the model. This customization enhances the model's ability to discriminate between healthy and pathological retinal regions, leading to more accurate ROP diagnosis.

    \item Voting System for Enhanced Decision-Making: To leverage the knowledge embedded in multiple retinal images of the same premature infant, we introduce a voting system. Each image contributes its prediction, and the final decision is made based on a consensus among the individual predictions. This approach helps mitigate the impact of variations in image quality, viewing angles, and potential misclassifications, thereby improving the overall accuracy and reliability of ROP detection.

\end{itemize}

Collectively, these contributions advance the field of ROP detection by addressing the challenges associated with limited real-world data, introducing a customized CNN model, and leveraging the collective knowledge of multiple images through a voting system. Our research provides valuable insights into improving the early detection and diagnosis of ROP, ultimately leading to timely interventions and better outcomes for premature infants at risk of vision loss.

\section{Related works}
Recent advancements in the field, exemplified by \cite{gensure2020artificial} and \cite{scruggs2020artificial}, have demonstrated the efficacy of Deep Learning (DL) in the early detection of ROP through retinal images. Notably, \cite{gensure2020artificial} highlights the transformative role of CNNs in ROP diagnosis, shedding light on the shift from conventional methods. This evolution is exemplified by successful models like DeepROP \cite{wang2018automated} and i-ROP DL \cite{brown2018automated}, underscoring both achievements and challenges in AI's practical implementation. The emphasis of these studies lies in validation, clinical integration, and regulatory implications.

In the research by Ding et al. \cite{ding2020retinopathy}, a hybrid technique for categorizing ROP stages is introduced. This method involves precise demarcation line identification through object segmentation, followed by integration as binary masks into original images. Subsequently, a CNN is deployed for stage classification. Despite its promise, the quality of segmentation and potential computational demands remain as considerations. Additionally, Mulay et al. \cite{mulay2019early} contribute significantly to ROP early detection using CNN-based techniques. While the paper highlights strengths, addressing the limitations could amplify its relevance in clinical practice.

Addressing ROP recognition, \cite{hu2018automated} introduces an innovative CNN architecture incorporating feature aggregation operators, enhancing robustness. Extensive evaluation using annotated examinations corroborates its effectiveness in multi-step recognition tasks and network performance analysis. However, data distribution imbalance, subjective annotation, and real-time feasibility pose challenges. To augment its impact, enhancing interpretability and extending applications to broader ophthalmological domains could be explored.

The structure of the rest of this paper is as follows: The "Materials and Methods" section delineates the foundational architecture and the rationale for adopting MobileNet and CNN. Following this, we present the Models. Subsequent to that, in the "Experiments" section, we undertake comprehensive evaluations of datasets, employing diverse preprocessing techniques. Finally, the "Conclusion, Discussion, And Future Works" section encapsulates our concluding remarks, discussions, recognition of limitations, and propositions for prospective research.

\section{Materials and Methods}
\label{sec: Method}
This section consist of dataset as materials of paper and then includes process of methods which started by preprocessing and continuances by providing a brief overview of CNNs and introducing a well-known pre-trained model called MobileNet. Subsequently, we present the design of our proposed customized CNN model. By leveraging the foundation of CNNs and incorporating our tailored modifications, we aim to enhance the model's performance and optimize its suitability for the specific task at hand. The detailed architecture and configuration of our customized CNN model will be elaborated upon, highlighting the unique features and improvements introduced to address the challenges associated with the ROP detection problem.

\begin{figure}[htp]

\begin{center}
 \includegraphics[width=1\textwidth]{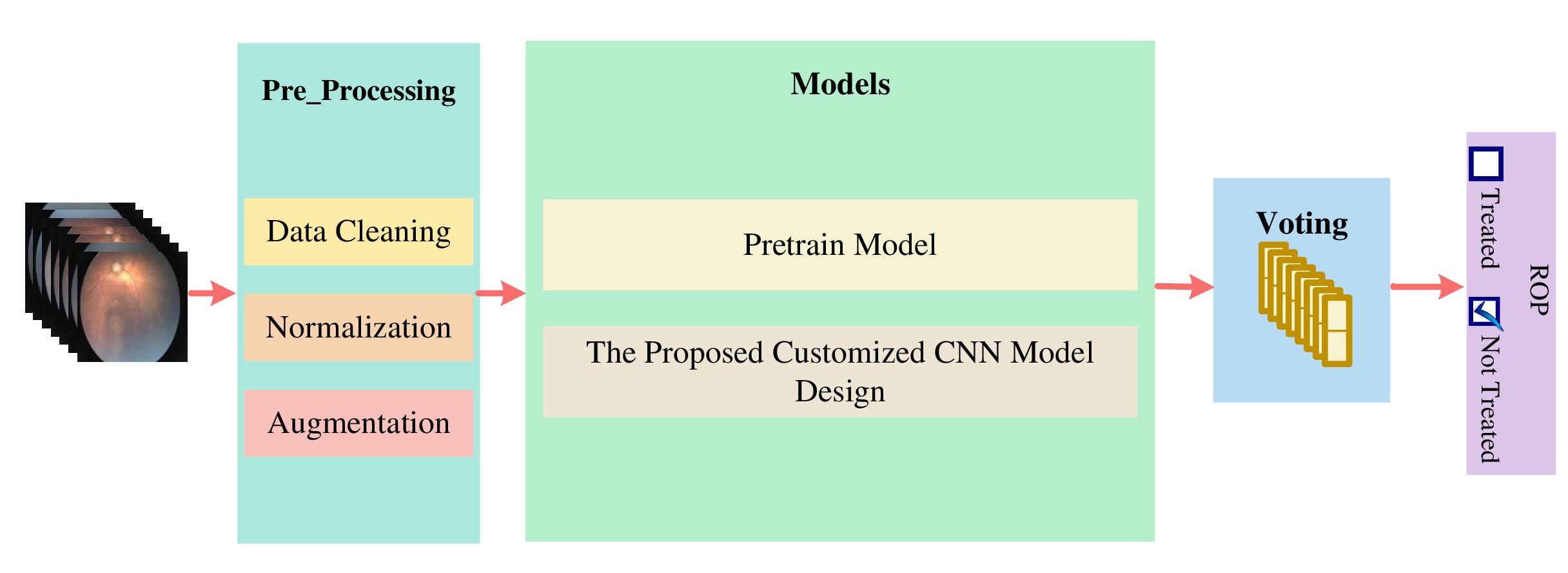}
\caption{Workflow Visualization for Automated Diagnosis of ROP using Customized CNN Models: Preprocessing, Feature Extraction, and Voting System Integration. Followed by preprocessing steps, the Augmented technique is utilized to alleviate imbalanced data challenges. The preprocessed images are then fed into the CNN models, enabling the extraction of crucial features. Our approach notably incorporates a voting system to improve ROP classification, effectively accounting for the varying angles captured in the retinal images.}
\label{fig:Flowchart}
\end{center}
\end{figure}

Figure \ref{fig:Flowchart} provides an overview of the paper's rationale, depicted in a flowchart. It highlights the key steps involved in pre-processing, model development, voting system integration, and the classification of ROP. The initial stage involves the preprocessing of retinal images acquired from various perspectives, aiming to optimize the data quality and prepare it for further analysis. Subsequently, the preprocessed images are fed into the customized CNN models, where sophisticated algorithms are applied to extract essential features and capture relevant patterns indicative of ROP presence. our approach incorporates a voting system that aggregates the predictions from multiple angles captured in the retina images. By considering the diverse perspectives and leveraging the collective decision-making power of the voting system.

\subsection{Dataset}
The research utilized authentic retinal images obtained from premature infants, which were captured from various angles. Multiple images were acquired for each infant, as each image encompassed approximately 170 degrees out of a total of 200 degrees of the eye. 
The dataset utilized in this study comprises two distinct image categories. The first category encompasses high-quality images characterized by appropriate color representation and dimensions of 480 × 640 pixels. Conversely, the second category consists of low-quality images with simplified color schemes and larger dimensions of 1200 × 1600 pixels.
For each infant, there are two categories of images corresponding to the left and right eyes. The number of images may vary for each category, but collectively, they cover the entirety of the eye with varying amounts of images. We obtained four sets of data, which are accessible through the provided link. These datasets were incorporated into the design and training process of the deep neural network model. This integration facilitated the exposure of the model to diverse data types and enabled the learning of dominant patterns within each class.

\subsection{Preprocessing}
\label{sec:preProcessing}

Several process is done on data which help to improve and prepare data for training in DL. Three of them are done for this section : 1-Data Cleaning, 2- Normalization, 3- Augmentation.

\subsubsection{Data Cleaning}
It is one of techniques of data science that analysis dataset which examine the label and data are related together or not. The data cleaning is consist of denoising, data selection and remove wrong or corrupted data.

\subsubsection{Normalization}
The range of values of dataset will be scale to specific range to prevent vanishing and exploding parameters value and help the model to process better. in this research, dataset includes of RGB images whose pixels have values between 0 and 255  and is normalized to the range between 0 and 1.
 
\subsubsection{Augmentation}

In many medical datasets, there is an imbalance between the number of ROP images and non-ROP images, with the former being fewer in quantity. To address this issue, data augmentation methods are commonly employed, as utilized in this research. Specifically, the dataset is subjected to various augmentation techniques that aim to preserve the features derived from the pixel values. These techniques include flipping the rows of pixels vertically or horizontally and rotating the images. By applying these operations, the dataset is augmented, resulting in increased diversity and size. Consequently, the performance and generalization ability of the classification model are enhanced.

In this study, various data augmentation techniques were utilized, including rotating the images by 90 degrees, flipping rows and columns, combining flipping and rotation, and enhancing contrast between image colors. These methods effectively expanded the dataset, resulting in a potential increase in the size. 

\subsection{Models}

The pretrained model and the design of the proposed customized CNN model are introduced in this section, and their corresponding code is accessible through the provided \href{https://github.com/DRAGON20-3/Retinopathy_of_Prematurity/tree/main/ROP_Detection}{GitHub Link}.

\label{sec:Models}
\subsubsection{Pretrain Model}
\label{sec:Pretrain}
\begin{itemize}
    
\item  MobileNet

MobileNet is a popular and influential architecture in the field of deep learning, specifically designed for efficient and lightweight image classification tasks on mobile and embedded devices.

We use the pre-train model of MobliNetV2 which classified 1000 class of images of ImageNet dataset according to convolution with input layer of 224$\times$224.

\end{itemize}

\subsubsection{The Proposed Customized CNN Model Design}
\label{sec:Costumize}
Convolutional Neural Networks (CNNs) are a class of deep learning models that are widely used in image classification tasks. These models consist of multiple layers, including convolution layers that apply filters to input images to extract relevant features. The trained neural network then performed classification tasks on the input data, determining whether each instance belonged to the ROP category or not.

While there are well-known CNN architectures such as MobileNet and ResNet that have been successful in image classification problems, we specifically developed a customized CNN model for ROP detection. This custom model was designed to outperform MobileNet in terms of accuracy and effectiveness in identifying ROP cases. By tailoring the architecture and incorporating domain-specific knowledge, we aimed to improve the detection performance and address the unique challenges associated with ROP diagnosis.

In our research, we have developed a customized CNN model specifically tailored for ROP detection. The design of this model incorporates several key components to optimize its performance and enhance its ability to accurately identify ROP cases.

One important aspect of the model design is the utilization of convolution layers at various depths within the network. These convolution layers are applied with different sizes and filter configurations, allowing the model to extract relevant features at multiple levels of abstraction. By employing this hierarchical approach, the model can capture both low-level details and high-level contextual information, enabling a more comprehensive analysis and representation of the input retinal images.

To address the challenge of gradient instability, normalization techniques have been integrated into the CNN model design. These normalization methods are applied to different layers of the network, each with varying dimensions. By normalizing the weights and activations, the model's performance is stabilized, mitigating issues such as vanishing or exploding gradients. This promotes more effective training and improves the overall accuracy of the model.

Through the careful design and incorporation of these components, our customized CNN model aims to outperform existing approaches, such as MobileNet, in ROP detection. By leveraging the strengths of CNNs and optimizing the network architecture, we strive to enhance the accuracy and reliability of ROP diagnosis, ultimately contributing to improved healthcare outcomes for premature infants.
    
\begin{figure}[ht]

\begin{center}
 \includegraphics[width=.80\textwidth]{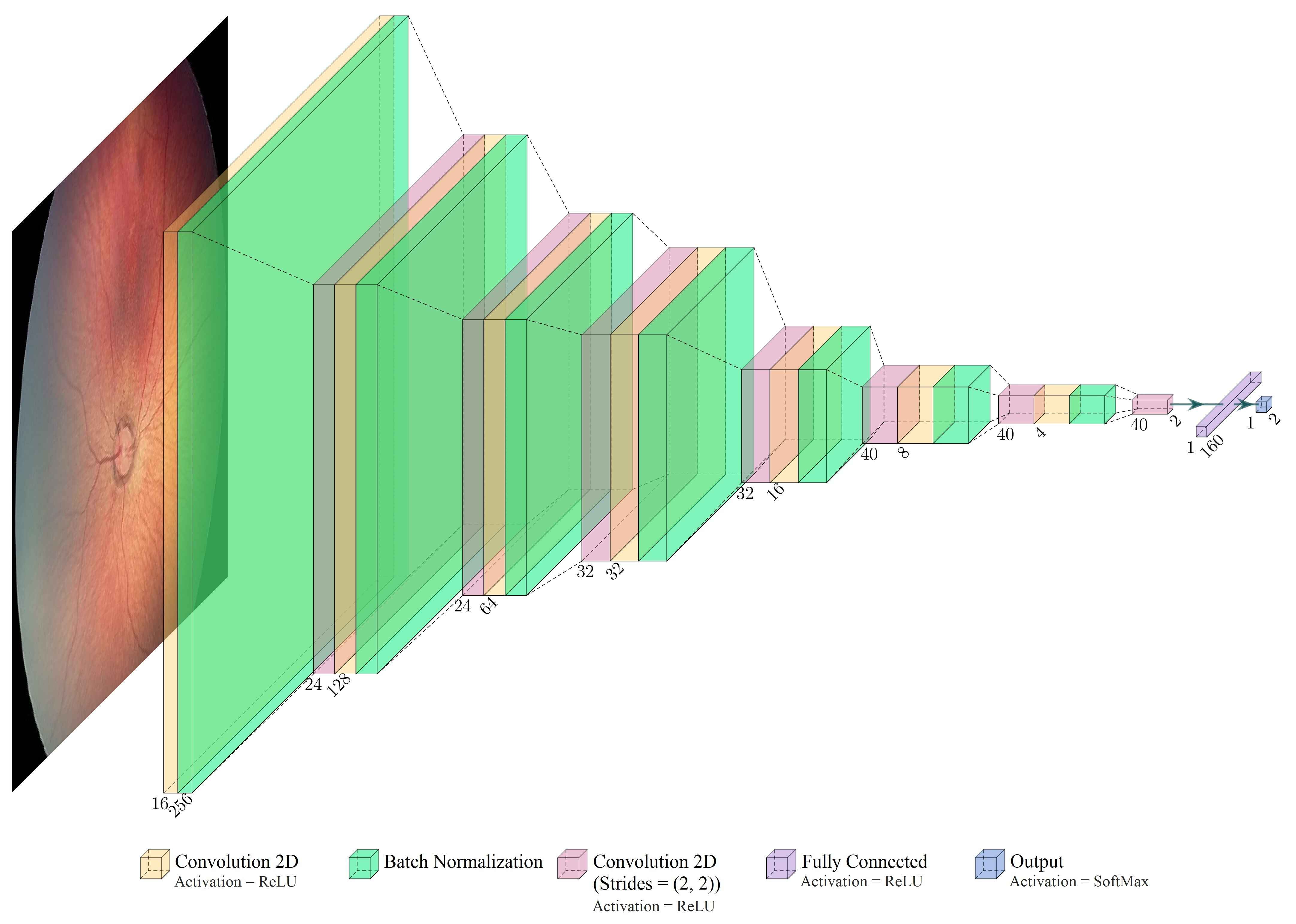}
\caption{The proposed CNN model architecture designed specifically for retinal image processing. It includes specialized components for effective feature extraction and classification. Convolution layers with varying stride sizes and ReLU activation functions are strategically incorporated to capture both local and global features. Batch Normalization is utilized to enhance model robustness against outliers. The fully connected layer acts as the last feature extractor layer, summarizing essential information from the retinal image. The model's output provides predictions for treatment decisions based on the extracted features.}
\label{fig:Proposed_Model}
\end{center}
\end{figure}

As illustrated in Figure \ref{fig:Proposed_Model} the Proposed Customized CNN Model, convolution layers are implemented with a stride of 1x1 and employ the ReLU activation function, visually represented by the yellow boxes. To improve the model's resilience against outliers, Batch Normalization is integrated into the architecture. The pink boxes correspond to convolution layers with a larger stride of 2x2 and ReLU activation function, allowing for the capture of more global features. Feature extraction from the retina image is accomplished through the utilization of a Fully Connected layer with ReLU activation function. Ultimately, the model's final output determines whether the patient should receive treatment or not based on the extracted features.

The loss function utilized in the introduced models, is Binary CrossEntropy. It is computed by considering the density function of both the ground truth and the prediction. The specific formulation is as follows:
\begin{equation}
    L(y,\hat{y})=y Log (\hat{y})+(1-y)Log (1-\hat{y})
\end{equation}

Where $y$ is ground truth and $\hat{y}$ is prediction of deep neural networks.

\section{Experimens}

\subsection{Research objectives}
\label{sec:Obj}
In this section, we present the experimental evaluation of the customized CNN model's performance. The experiments are designed to address the following three research objectives:

\begin{itemize}
    \item \textbf{Research Objective 1: Accuracy of Models} Our analysis focuses on evaluating the impact of each preprocessing technique on both the pretrain and customize CNN models to achieve efficient automatic diagnosis of ROP.
    
    \item \textbf{Research Objective 2: Cost of Models} We investigate the computational cost of each model and explore the coherence of customize CNN models leading to improved accuracy, as discussed in Research Objective 1.
    
    \item \textbf{Research Objective 3: Time Complexity of Models} Considering the need for efficient automatic diagnosis of ROP in real-world systems, we examine the time complexity of the methods.
    
\end{itemize}

Through these experiments, we aim to gain insights into the model's capabilities and effectiveness in various scenarios
\subsection{Experimental setup}

The experimental setup for this study involved the implementation of the proposed methodology using the Python programming language and the TensorFlow package. The computations were performed on a system equipped with a GPU T4, a Intel Xeon CPU with 2 vCPUs (virtual CPUs), and 12 GB of RAM. The process consisted of several steps, including data preparation, dataset splitting, data augmentation, model design, training, performance estimation, and result analysis. Initially, the images and their corresponding labels were read, followed by preprocessing to ensure compatibility with the model, The images are resized to match the input size required by the model. The dataset was then divided into train and test sets, allowing for proper evaluation, the training set consists of 80\% of all the images, although in some cases, this percentage may be increased to prevent overfitting, which is a phenomenon where the model becomes too specialized to the training data and performs poorly on unseen data. Data augmentation techniques were applied to enhance the diversity of the training data. Subsequently, an appropriate neural model was designed and configured, taking into account the specific requirements of the task. The model was trained on the training dataset, optimizing its parameters iteratively. Performance metrics were calculated to evaluate the model's performance on both the train and test datasets. Finally, a comprehensive analysis of the results was conducted to gain insights into the model's effectiveness and identify potential areas for improvement.

The optimizer employed in our model is Adaptive Moment Estimation (Adam) with a learning rate of $10^{-3}$. Additionally, training is conducted with 100 epochs, and the obtained results are reported accordingly.

\subsection{Results}

There are three research objective in this section which analyse models that said in \ref{sec:Models}.

\subsubsection{Research Objective 1: Accuracy of Models}

In this section we show the result of effectiveness of preprocessing option which said in \ref{sec:preProcessing} to evaluate the classification of both pretrain and customized CNN models that said in  \ref{sec:Pretrain}

\begin{itemize}
    \item Pretrain model Performance
\end{itemize}
\label{sec: EXPMobileNet}
In this experimental section, the performance of the pretrain model on the dataset is examined. The evaluation involves implementing various preprocessing tools separately to assess the effectiveness of each item, as depicted in Figure \ref{fig:Flowchart}. 
Table \ref{tab:table1} provides an overview of the various preprocessing techniques used for the pretrain MobileNet network. The first technique employed is normalization, which rescales the pixel values of each image to a range of zero to one. By doing so, all images in the dataset are aligned to have consistent pixel value ranges, these values are more numerically stable. Large pixel values can lead to numerical overflow or underflow issues, especially when using certain activation functions or optimization algorithms.

In Table \ref{tab:table1}, the first and second rows contain 80\% of the data for training and 20\% for testing. Subsequently, in the following steps, the proportion of test data is increased to 30\%. This increase in test data allows for a more reliable assessment of the model's performance. Moreover, it ensures that the model is not simply memorizing the training data but genuinely learning the underlying patterns.

The subsequent improvements that elevate accuracy from 66.86\% to 69.71\% rely on augmenting the dataset with additional data. Nevertheless, this data augmentation process introduces increased computational demands and poses a risk of overfitting in case of model complexity or noisy data. On a positive note, the enlarged dataset facilitates superior generalization to novel, unseen data and mitigates bias by offering a more comprehensive and diverse collection of examples.

The subsequent column, denoted as Conditional Training, represents a technique utilized to enhance accuracy and optimize MobileNet's parameter settings. The outcomes presented in Table \ref{tab:table1} demonstrate that conditional training initially results in a reduction of accuracy and F-score. Nonetheless, it is noteworthy that this step contributes to a subsequent increase in accuracy beyond the previous results.

Next, we explore the impact of the augmentation techniques mentioned in \ref{sec:preProcessing} on the pre-trained MobileNet model. The results demonstrate that this option leads to a significant increase in accuracy and F-score, with improvements of 12\% and 0.12, respectively. The dataset was manually cleaned to eliminate incorrect data points containing wrong labels or images with wrong features and the result show on table in finally row of table \ref{tab:table1}. Dirty data, characterized by the presence of noise and outliers, can potentially confuse machine learning models and impact their performance. However, there are situations where data cleaning may not yield substantial improvements or prove as beneficial. Aggressive data cleaning can lead to the removal of valuable information, which might be crucial for the learning process of the model. Excessive data cleaning may also result in the loss of diversity and useful patterns.

\begingroup
\setlength{\tabcolsep}{5 pt} 
\renewcommand{\arraystretch}{1.9} 
\begin{table}[htp]
\caption{Performance criteria of the pretrain MobileNet method}
\fontsize{9pt}{9pt}\selectfont
\label{tab:table1}
\begin{tabular}{lccccccccc}
\hline
Method & Normalization & \makecell[l]{Increase \\ Test \\Data}  & \makecell[l]{Increase \\ Data}  & \makecell[l]{Conditional \\ Training}   & Augmentation & \makecell[l]{Clean \\ Data}  & \makecell[l]{Number of \\Data} & Accuracy & F1-Score \\ \hline
\multirow{7}{*} {MobileNet} &           &           &           &           &           &           & 1045 & 72.57 & 0.73 \\
 & $\checkmark$ &           &        &           &           &           & 1045 & 60.57 & 0.61 \\
 & $\checkmark$ & $\checkmark$ &           &           &           &           & 1045 & 66.86 & 0.67 \\
 & $\checkmark$ & $\checkmark$ & $\checkmark$ &          &           &           & 1256 & 69.71 & 0.70 \\
 & $\checkmark$ & $\checkmark$ & $\checkmark$ & $\checkmark$ &           &           & 1265 & 60.00 & 0.60 \\
 & $\checkmark$ & $\checkmark$ & $\checkmark$ & $\checkmark$ & $\checkmark$ &           & 1256 & \textbf{72.00} & \textbf{0.72} \\
 & $\checkmark$ & $\checkmark$ & $\checkmark$ & $\checkmark$ & $\checkmark$ & $\checkmark$ & 1022 & 65.71 & 0.66 \\ \hline
\end{tabular}
\end{table}
\endgroup
\begin{itemize}
    \item Customized CNN model performance
\end{itemize}
\label{sec:EXPcostum}
In this section, we conduct an analysis of the Customized CNN model's performance on the dataset. The evaluation process entails the individual implementation of various preprocessing techniques to assess the effectiveness of each approach, as depicted in Figure \ref{fig:Flowchart}. An overview of the different preprocessing techniques and various model architecture utilized for the Customized CNN model is presented in Table \ref{tab:table2}. 

All the preprocessing steps mentioned in Table \ref{tab:table1} have been implemented and are presented in Table \ref{tab:table2}. In Table \ref{tab:table2}, the initial preprocessing technique utilized is Partial Training, which aims to address challenges related to unbalanced data and low RAM capacity in machine learning. This method involves the iterative training of the model on smaller subsets, known as mini-batches, instead of using the entire dataset at once. Nevertheless, as indicated in the table, Partial Training might lead to a performance reduction in the model. When only a subset of the data is used for training, the model may not fully leverage the diversity present in the complete dataset, potentially resulting in suboptimal performance.
As shown in row two and three of Table \ref{tab:table2}, increasing the dataset size is one of the key points in the experimental phase that contributes to the enhancement of Accuracy. This is because the model can better comprehend the dominant patterns present in various images when exposed to a larger and more diverse dataset. It is worth noting that some images possess low quality due to their camera sources, and it is necessary to include such images in the training process, as they are representative of real-world scenarios. In row four of  Table \ref{tab:table2}, when we had 1736 images that comprised both low and high-quality data, we modified the probability of selecting data points, where the low-quality data had twice the chance of being selected compared to high-quality data. This technique helps the model to effectively learn from low-quality data, effectively incorporating it into the learning process.
Before modifying the probability of selected data with 249,778 parameters, the experimental phase involved a step of decreasing the convolution channels. In row six of Table \ref{tab:table2}, the number of parameters was reduced to 124,714, resulting in an increase in accuracy from 62.86 to 73.14. However, in the subsequent row, further decreasing the convolution channels to 65,554 resulted in a decrease in accuracy from 73.14 to 68.57. As a result, the model with 124,714 parameters was chosen and subsequently fine-tuned.

\begingroup
\setlength{\tabcolsep}{0.5 pt} 
\renewcommand{\arraystretch}{1.9} 
\begin{table}[htp]
\caption{Performance criteria of the Proposed Customize method}
\fontsize{9pt}{9pt}\selectfont
\label{tab:table2}

\begin{tabular}{lccccccccccc}
\hline
Method & \makecell[l]{Partial \\ Training} & \makecell[l]{Increase \\ Data}  & \makecell[l]{Increase \\ Low quality \\ Data}  & \makecell[l]{Change \\ Probability of \\ Selected \\Data}   & \makecell[l]{Decrease \\ Convolution \\ Channels} & \makecell[l]{2$x$ Decrease \\ Convolution \\ Channels}  & \makecell[l]{Fine-tune} & Voting & \makecell[l]{Number of \\Data} & Accuracy & \makecell[l]{F1-Score}  \\  \hline
\multirow{8}{*} {\makecell[l]{Customize \\ CNN} } 
 & $\checkmark$  &           &           &           &           &          & & & 1022 & 63.43 & 0.63 \\
 & $\checkmark$ & $\checkmark$          &        &           &           &          && & 1256 & 74.29 & 0.74 \\
 & $\checkmark$ & $\checkmark$ & $\checkmark$          &           &           &          & && 1390 & 62.86 & 0.63 \\
 & $\checkmark$ & $\checkmark$ & $\checkmark$ & $\checkmark$         &           &        & &  & 1736 & 62.86 & 0.63 \\
 & $\checkmark$ & $\checkmark$ & $\checkmark$ & $\checkmark$ & $\checkmark$          &       &   & & 1736 & 73.14 & 0.73 \\
 & $\checkmark$ & $\checkmark$ & $\checkmark$ & $\checkmark$ & $\checkmark$ & $\checkmark$      &  &  & 1736 & 68.57 & 0.69 \\
 & $\checkmark$ & $\checkmark$ & $\checkmark$ & $\checkmark$ & $\checkmark$ &   & $\checkmark$ & &1911 & 89.71 & 0.90 \\ 
  & $\checkmark$ & $\checkmark$ & $\checkmark$ & $\checkmark$ & $\checkmark$ &   & $\checkmark$ & $\checkmark$ & 1911 & \textbf{99.99} & \textbf{0.99} \\\hline
\end{tabular}
\end{table}
\endgroup

During the fine-tuning phase, the model is trained using the entire dataset, which comprises 1911 images from both the training and validation sets. Finally, we utilize the voting technique described in Section \ref{sec: Method}, leading to a remarkable final accuracy of 99.99\%. 
\subsubsection{Research Objective 2: Cost of Models}
The primary goal of this section is to evaluate the model's performance in relation to its computational cost. To achieve this objective, the accuracy and loss function flowchart are depicted in Figure \ref{fig:figure_chart}. Figures \ref{fig:figure_chart} (a), (c), and (e) depict the accuracy of models in each iteration, while Figures \ref{fig:figure_chart} (b), (d), and (f) show the binary cross-entropy loss function of models in each iteration. The blue lines in these figures represent the accuracy or loss of the training data, and the red lines represent the accuracy or loss of the validation data.

Figures (a) and (b) in Figure \ref{fig:figure_chart} correspond to the best MobileNet model discussed in Table \ref{tab:table1} of Section \ref{sec: EXPMobileNet}. Figures (c) and (d) are associated with the pre-fine-tuned Customized CNN model mentioned in row 5 of Table \ref{tab:table2} in Section \ref{sec:EXPcostum}. Finally, Figures (e) and (f) are related to the fine-tuned Customized CNN model stated in row 7 of Table \ref{tab:table2} in Section \ref{sec:EXPcostum}. 

In Figures (a), (c), and (e), the close proximity of the train line and the validation line suggests that the model performs well on both the training and validation data. This indicates that the model has successfully learned to capture the underlying patterns and features of the data without merely memorizing specific examples. On the other hand, the loss function plots in Figures (b), (d), and (f) represent the model's performance during training, and its value should be minimized as the training progresses. However, if there is a significant gap between the train line and the validation line, it might indicate that the model is likely capturing noise and specific details from the training data rather than truly grasping the fundamental patterns.

\begin{figure}[!t]
\centering

\includegraphics[width=0.5\linewidth]{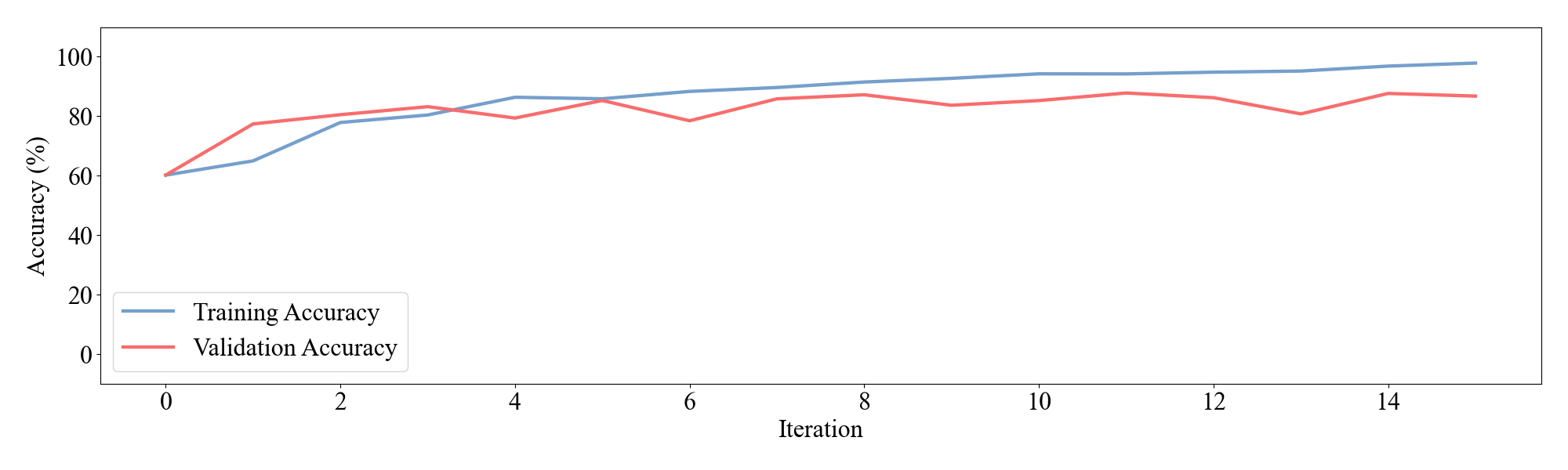}\\
{(a) Accuracy of the best MobileNet}

\vspace{12pt}
\includegraphics[width=0.5\linewidth]{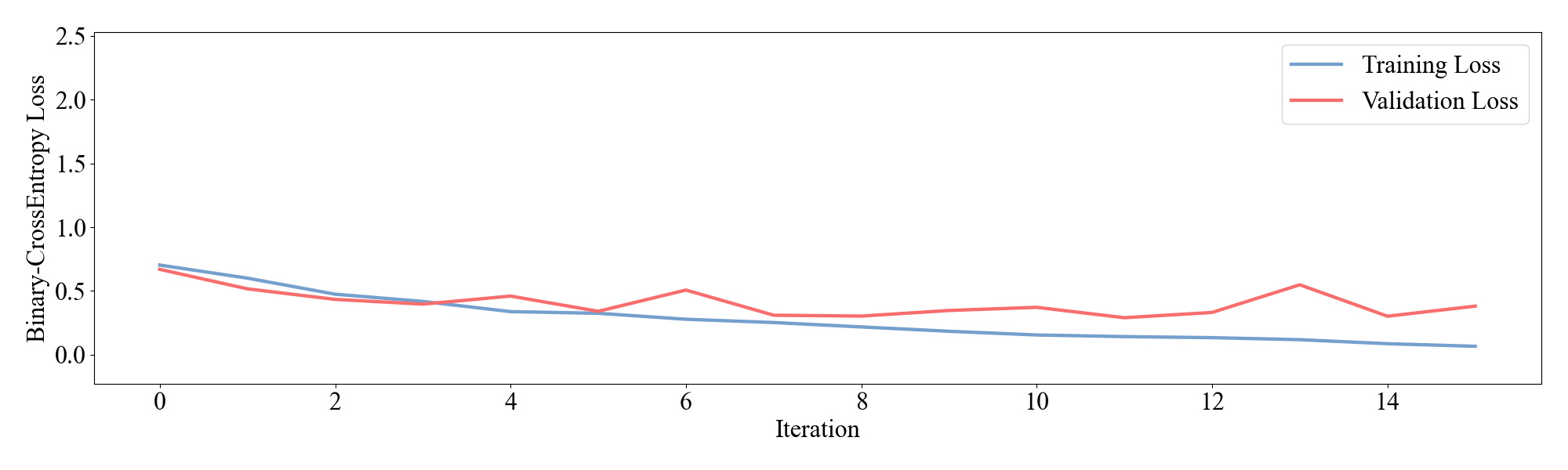}\\
{(b) Loss of the best MobileNet}

\vspace{12pt}
\includegraphics[width=0.5\linewidth]{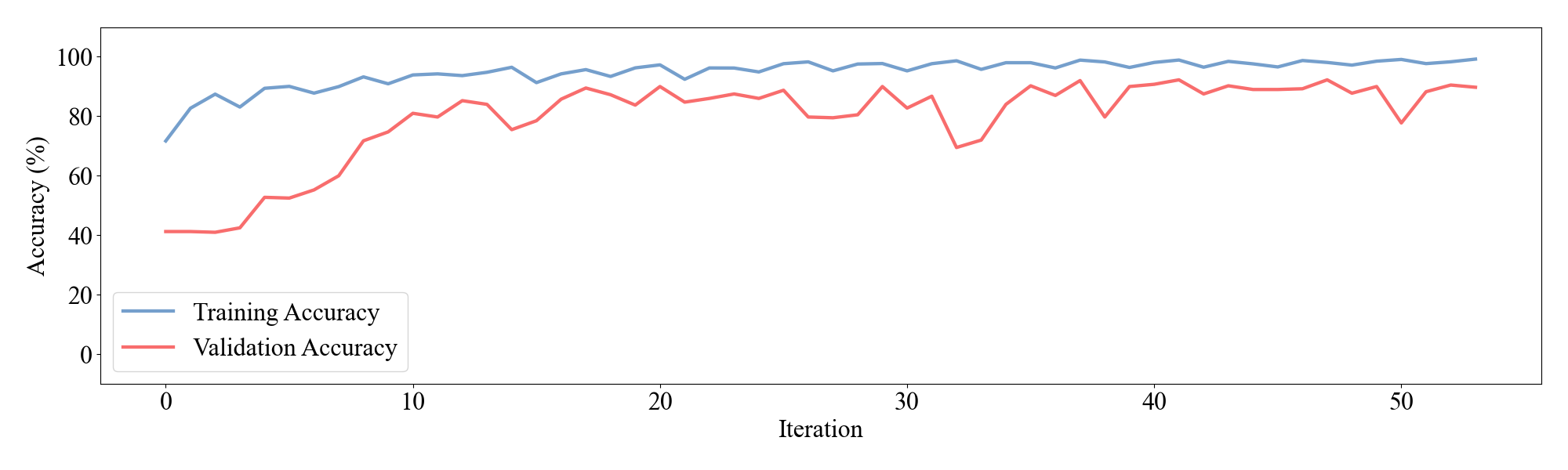}\\
{(c) Accuracy of pre-fine-tune Customized CNN}

\vspace{12pt}
\includegraphics[width=0.5\linewidth]{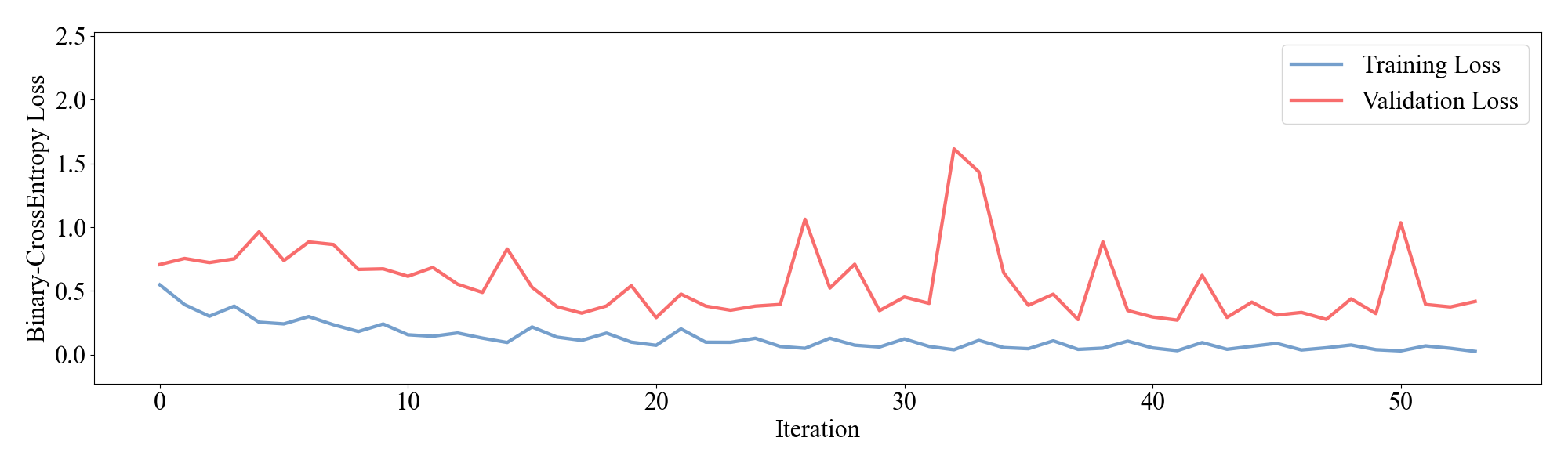}\\
{(d) Loss of pre-fine-tune Customized CNN}

\vspace{12pt}
\includegraphics[width=0.5\linewidth]{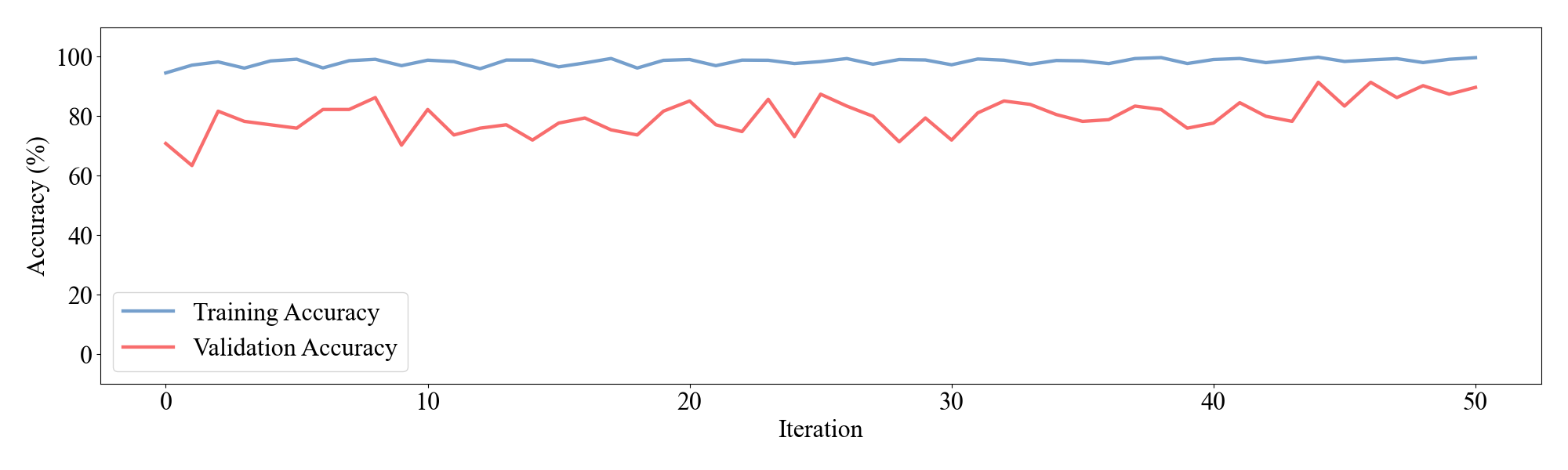}\\
{(e) Accuracy of fine-tuned Customized CNN}

\vspace{12pt}
\includegraphics[width=0.5\linewidth]{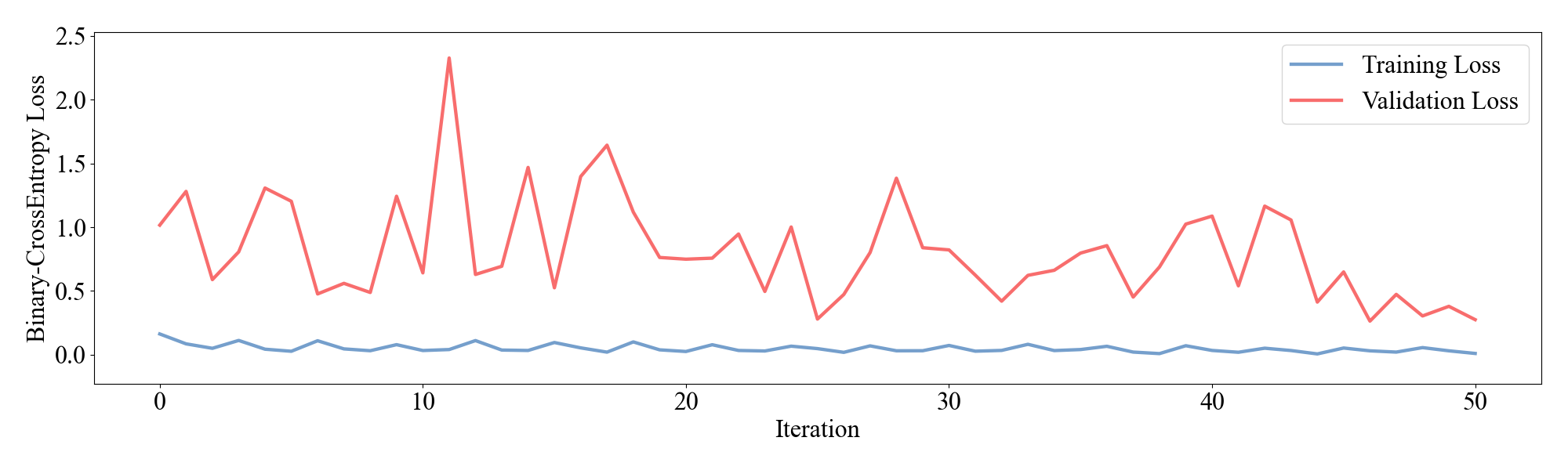}\\
{(f) Loss of fine-tuned Customized CNN}

\caption{Training and validation curves for the three best models. 
(a) Accuracy and (b) loss of the best MobileNet; 
(c) accuracy and (d) loss of Customized CNN before fine-tuning; 
(e) accuracy and (f) loss of Customized CNN after fine-tuning. 
Blue lines represent training data; red/orange lines represent validation data.}
\label{fig:figure_chart}

\end{figure}

\subsubsection{Research Objective 3: Time complexity of Models}

In this section, we aim to analyze the time complexity of the pretrain MobileNet and customized CNN models by computing the Frames Per Second (FPS). FPS is a measure of the number of frames processed by the models in one second. By calculating the FPS, we can estimate the time required for these models to process a given amount of data. This analysis will provide insights into the computational efficiency and performance of the two models.

Figure \ref{fig:figure_chart} (a) and (b) display the calculation of frames per second (FPS) for images processed through three models format: Tensorflow (TF), Tensorflow Lite with invocation (TFLite + Invoke), and Tensorflow Lite without invocation. The Tensorflow (TF) model represents the standard processing using the Tensorflow framework, while the Tensorflow Lite with invocation (TFLite + Invoke) model signifies the usage of Tensorflow Lite with additional invocation methods. On the other hand, the Tensorflow Lite without invocation model demonstrates the performance of Tensorflow Lite without any invoking procedures.

Figure \ref{fig:figure_chart} (a) demonstrates the runtime for 46 images using two different runtime: 10 and 100, for both MobileNet and Customized CNN models. In part (b), we calculate the runtime for a random number of images with the same runtime. The FPS values are then normalized and presented in the bar chart.  The pretrain MobileNet model achieves a higher FPS than the Customize CNN model due to the increased computation complexity of different architecture of convolution which MobileNet using 1$\times$1 filter size. However, despite the lower FPS, the Customize CNN model demonstrates better accuracy compared to the pretrain MobileNet model. It is noteworthy that the FPS of both models with 100 runtimes is especially crucial for real-world deployment, particularly when images are randomly selected.

\begin{center}
\begin{tabular}[htp]{cccccc}

\includegraphics[width=0.49\linewidth]{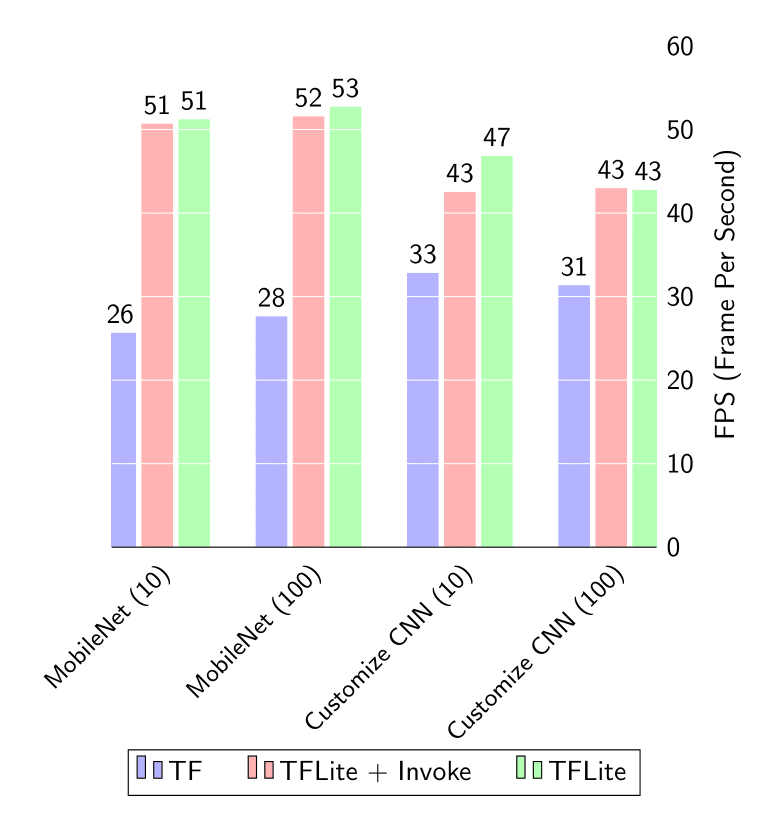} &
\includegraphics[width=0.49\linewidth]{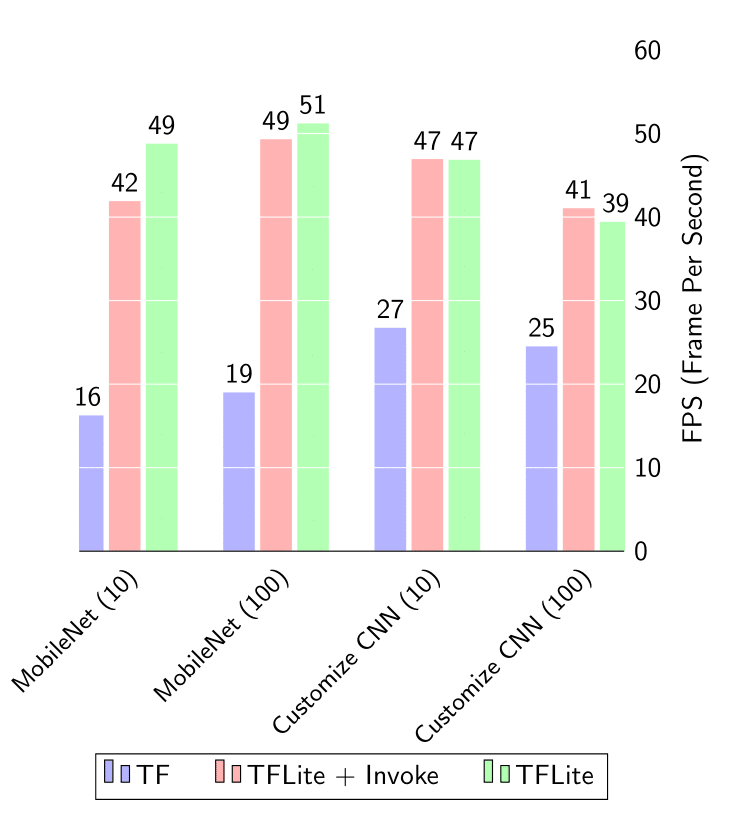} \\
 a) Selected 46 images & b) Selected random amount of images 
 \end{tabular}
\captionof{figure}{Time complexity analysis of pretrain MobileNet and customized CNN models using Frames Per Second (FPS) computation. Part (a) shows FPS for 46 images with runtimes of 10 and 100 for both models. Part (b) presents runtime for random images.  }
\label{fig:figure_chart}
\end{center}

\section{Conclusion, Discussion, And Future Works}

ROP is a retinal disorder that predominantly impacts premature infants, leading to considerable difficulties in its early diagnosis and management. To tackle these challenges and enhance the accuracy of ROP screening, the research introduces and evaluates several screening methods. The findings from this research provide insights into the effectiveness and potential improvements of these methods.

Several AI methodologies have been introduced to facilitate the diagnosis of ROP. These methodologies encompass diverse approaches, including the utilization of traditional machine learning techniques \cite{wittenberg2012computer} as well as deep learning models \cite{tong2020automated, mulay2019early, hu2018automated, ding2020retinopathy}.
The process of employing AI models to diagnose ROP through retinal images typically involves a series of stages, including dataset selection, preprocessing, feature extraction and selection, and finally, classification.

A total of 1911 authentic retina images were collected from premature infants at Khatam-al-Anbia Eye Hospital. The dataset is categorized into two distinct groups of images. The first category comprises high-quality images with accurate color representation, while the second category includes low-quality images with simplified color schemes. During the preprocessing phase, several steps were undertaken. Firstly, data cleaning was performed to ensure that labels and data were appropriately aligned. Next, normalization was carried out to scale the dataset values within a specified range. Additionally, augmentation techniques were applied to address the imbalance between the number of images depicting retinopathy of prematurity (ROP) and those without ROP.

The pre-trained MobileNet neural network was subjected to various preprocessing techniques on the dataset, including normalization, diverse augmentations, and cleaning. In this section, the optimal model achieved an accuracy of 72.00\%, which stands as the highest accuracy among all the employed preprocessing methods.

In the subsequent sections, deep learning (DL) techniques for diagnosing ROP using retinal images are employed. The customized CNN architecture consists of three essential elements: 1-Convolution layers with a stride of 1x1 and the ReLU activation function are implemented. 2- Batch Normalization is integrated into the architecture to enhance the model's robustness against outliers. 3- Convolution layers with a larger stride of 2x2 and the ReLU activation function are utilized. In the proposed customize CNN networks, when the stride of convolution filters increases, the features tend to become more generalized rather than detailed. Eventually, 160 features are extracted from each image using this configuration, and the model proceeds to diagnose ROP.


In Section \ref{sec:Obj}, we introduced three distinct research objectives. These objectives were formulated to assess the efficacy of the customized CNN model, examine the computational cost of the models in relation to the dataset, and investigate the influence of time complexity on systems. Concerning the first research objective, it was observed that the customized CNN model outperforms a significant majority of pretrain MobileNet and CNN models in terms of accuracy and F1-score within the dataset. Additionally, the integration of a voting system led to an approximate 10\% increase in accuracy and an enhancement of the F1-score by 0.9.

In regard to the second research objective, our experiments demonstrate that the incorporation of the proposed customized CNN model can substantially reduce the costs associated with deep neural network models. Lastly, concerning the third research objective, the outcomes of our experiments indicate that the suggested customized CNN models can be effectively deployed on specialized software and hardware platforms. This deployment can significantly contribute to efficient ROP diagnosis via retina images and may also serve as an auxiliary diagnostic tool in medical facilities.

In the subsequent sections, potential avenues for future research in the domain of ROP diagnosis through retina images are explored. One notable direction is the refinement and expansion of ROP diagnosis to encompass specific and detailed aspects such as diagnosing the stage and zone of the disorder. Various scholars have highlighted the efficacy of CNN models in effectively diagnosing neural disorders through retina images \cite{wittenberg2012computer, hu2018automated}.

As highlighted in the section discussing the limitations of this study, the dataset employed here is primarily focused on diagnosing ROP disorder. However, a significant scope for future investigations lies in compiling ROP datasets that account for various scenarios, zones, and stages of the disorder. One prospective research endeavor involves constructing classification models based on deep learning techniques, tailored for distinct zones and stages. This pursuit necessitates the availability of new labeling protocols for retina images. delving into more specific diagnostic details, leveraging advanced neural network models, and extending dataset diversity could greatly enrich the field of ROP diagnosis via retina images.

\bibliography{ROP}
\end{document}